\documentstyle[11pt,paspconf,epsf,amssymb]{article}
 
\begin{document}
 
\title{ An Automated Search for Metal-Poor Halo Stars in the
  Hamburg/ESO Objective-Prism Survey}
 
\author{Norbert Christlieb\altaffilmark{1}, Lutz Wisotzki \& Dieter Reimers}
\affil{Hamburger Sternwarte, Germany}
\author{Thomas Gehren \& Johannes Reetz}
\affil{Universit\"ats-Sternwarte M\"unchen, Germany}
\author{Timothy C. Beers}
\affil{Department of Physics \& Astronomy, Michigan State University, U.S.A.}
\altaffiltext{1}{E-mail: nchristlieb@hs.uni-hamburg.de}
 
\begin{abstract}
An automated search for metal-poor stars is carried out in the course of the
Hamburg/ESO objective-prism survey (HES), which covers the full southern sky
at Galactic latitudes $|b|\gtrsim 30\deg$.  As the HES reaches $\sim 1$
magnitude deeper and covers areas of the sky which have not been touched by
the HK Survey of Beers and collaborators, the total survey volume of the HES
represents an increase by a factor of 4.5 compared to the HK Survey.
Because of the limited availability of telescope time for spectroscopic
follow-up observations, we will focus on a search for the most metal-poor,
unevolved stars. We present a simulation study of the HES selection function
$P(B_J, T_{\mbox{\scriptsize eff}}, [\mbox{Fe/H}], \log g)$ for metal-poor
stars, and results from a first follow-up campaign at the ESO-NTT.
\end{abstract}
 
 
\section{Introduction}
 
Metal-poor stars have atmospheres that consist of the most primitive
(luminous) matter in the Universe. They offer the possibility to study the
formation and early chemical evolution of our Galaxy, and the first epochs of
star formation. Moreover, they are of significance for observational
cosmology, e.g., by setting a lower limit to the age of the Universe by their
ages, and by measurement of the abundances of light element lithium, thought
to have been formed in Big Bang nucleosynthesis.
 
As a result of the interference-filter/objective-prism survey of Beers and
collaborators (Beers et al. 1992; hereafter BPS92), the so-called ``HK
Survey,'' the number of known metal-deficient stars has increased dramatically
in the last few years.  However, as emphasized by Beers (1999, this volume),
many interesting projects to be undertaken in the near future require still
larger samples of {\em extremely\/} ($[$Fe/H$]<-2.5$) metal-poor stars.
 
Given this fact, but also facing the limited availability of observing time
for spectroscopic follow-up observations, we have decided to focus on a
sub-sample of the metal-poor stars being present on our plates, namely the
most metal-poor, unevolved stars (main-sequence stars, turnoff-stars and
subgiants). The restriction of the candidate set to these stars results in a
tractable number of candidates per field. In section 3 we show how the
selection can be optimized towards this aim.
 
With the advent of large multi-object spectrographs, e.g., 6dF (Watson 1998),
projects involving follow-up observations of a larger number of metal-poor
stars should soon be feasable, such as detailed investigation of the
metallicity distribution function of the Galactic Halo (Beers 1999, this
volume).
 
\section{ The Hamburg/ESO Survey}
 
The HES (Wisotzki et al. 1996) is an objective-prism survey primarily
targetted for bright QSOs, covering the full southern extragalactic sky ($\sim
10\,000$~deg$^2$). It is based on IIIa-J plates which have been taken with the
ESO Schmidt telescope and its 4$^{\circ}$ prism, yielding a wavelength range
of $3200\,\mbox{\AA} < \lambda < 5300\,\mbox{\AA}$ and a seeing-limited 
spectral resolution of 15\,{\AA} at H$\gamma$.
 
So far 80\,\% of the plates have been digitised in Hamburg using a PDS~1010G
microdensitometer. Scanning and reduction of all plates is expected to be
completed by the end of 1998. Apart from the quasar survey work, this data
base of digitised moderate-resolution spectra opens the possibility of a
fast and efficient exploitation of the {\em stellar\/} content of the survey,
which includes, among many other interesting classes of objects, the most
metal-poor stars.
 
\begin{figure}[htb]
  \plotfiddle{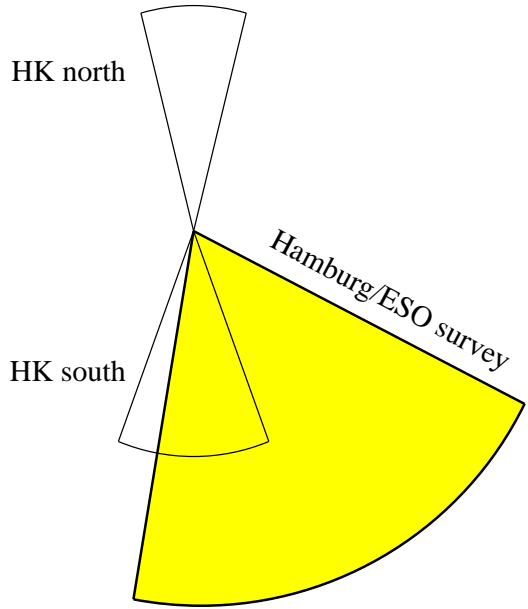}{7.5cm}{0}{100.}{100.}{-102}{0}
  \caption{\label{surveycompare}Comparison of HK survey
    and Hamburg/ESO survey volumes, assuming limiting magnitudes of
    $B_J=15.5$ and $B_J=16.5$, respectively. By the HES the total
    survey volume for metal-poor stars can be increased by a { factor
    of 4.5} compared to the HK survey alone. 
    }
\end{figure}

Because the limiting magnitude of the HES plates ($B\sim 16.5$) is about one
magnitude fainter than that of the HK Survey ($B\sim 15.5$), and a large area
in the southern extragalactic sky which is not covered by the HK survey is
covered by the HES (cf. Fig. \ref{surveycompare}), the HES has the potential
for another significant increase of the number of known metal-poor stars.
 
\section{Selection of Candidates}\label{canselect}
 
As a benchmark for a selection of metal-poor stars by automatic
classification, a technique which is under development (Christlieb et
al. 1997; 1998a,b), we implemented a search for spectra exhibiting a Ca\,{\sc
ii}~K line weaker than normal for its colour. Thanks to the $UBV$ photometry
of the HK survey, it was possible to perform a calibration between $B-V$ and
the half-power point of the density distribution of the HES spectra in the
range 3900\,{\AA}$<\lambda <$5300\,{\AA} (named \texttt{x\_hpp2} in the
following). The r.m.s. error of the calibration is $0.1$\,mag, more than
sufficient for this purpose.
 
``Normal'' Ca\,{\sc ii}~K line strengths are defined plate-by-plate by a
straight line fit of the Ca\,{\sc ii}~K index as a function of
\texttt{x\_hpp2}. The fit is performed for all point-like, overlap-free
sources with $S/N>15$ on one plate, after outliers (e.g., metal-poor stars,
but also many plate artifacts) have been removed by an iterative
$\kappa\sigma$-clipping.
 
\subsection{A Simulation Study of the Metal-Poor Star Selection Function}
 
In a first simulation study we investigated the composition of a sample of all
objects $3\sigma$ below the Ca\,{\sc ii}~K index fit. For this we used a grid
of model spectra with metallicities between $[$Fe/H$]=-1.0$ and $-3.0$ (Tab.
\ref{selectionfunction}).  The model spectra have been converted to artificial
objective-prism spectra by rebinning and multiplication with spectral
sensitivity curves of the HES objective-prism plates. From Tab.
\ref{selectionfunction} one can see that above $S/N=15$ (corresponding to
$B_J<16.0\dots 16.5$, depending on plate quality) there is a fairly sharp
cutoff between $[$Fe/H$]=-1.0$ and $[$Fe/H$]=-2.0$, and the simulated samples
of stars with $[$Fe/H$]\le -2.0$ have a completeness close to 100\,\%. The
restriction of the sample to lower metallicities is limited by the spectral
resolution of the HES objective-prism spectra: For stars in the considered
temperature range, Ca\,{\sc ii}~K reaches the detection limit at $[$Fe/H$]\sim
-2.0$.
 
{\small
\begin{table}
\caption{\label{selectionfunction} Selection function in dependence of
  stellar parameters $T_{\mbox{\scriptsize eff}}$, $\log g$, and $[$Fe/H$]$, at
  different magnitudes, for a selection by \texttt{CaKindex}$<3\sigma$ below
  linear fit of \texttt{CaKindex} on one randomly choosen plate. For each
  class, 300 simulated model spectra have been used.}
  \begin{center}
    \begin{tabular}{cccc|rrrr|rr}\tableline\tableline
  \rule{0.0mm}{2.5ex} & & & &\multicolumn{6}{c}{\% selected on randomly
    choosen plate}\\
  $T_{\mbox{\scriptsize eff}}$ & $\log g$ & $[$Fe/H$]$ & $B-V$ &
  \multicolumn{1}{c}{$S/N=30$} & \multicolumn{1}{c}{25} &
  \multicolumn{1}{c}{20} & \multicolumn{1}{c|}{15} &
  \multicolumn{1}{c}{10} & \multicolumn{1}{c}{5}\\ 
  & & & & \multicolumn{1}{c}{$B_J=14.7$} & \multicolumn{1}{c}{15.2} &
  \multicolumn{1}{c}{15.7} & \multicolumn{1}{c|}{16.1} &
  \multicolumn{1}{c}{16.7} & \multicolumn{1}{c}{17.3} \\\tableline 
  6000 & 4.50 & -1.00 & 0.50 & 0.0 &   0.0 &   0.0 &   0.0 &  20.0
  & 37.3\rule{0.0mm}{2.5ex}\\

  6000 & 4.50 & -2.00 & 0.48 & 98.3 &  96.7 &  91.3 &  81.0 &  98.7 &  84.3\\

  6000 & 4.50 & -3.00 & 0.47 &  100.0 & 100.0 & 99.7 & 99.7 & 100.0 & 93.7\\[0.5ex]

  6500 & 3.50 & -1.00 & 0.39 & 0.0 &   0.0 &   0.0 &   0.0 &  43.3 &  43.0\\

  6500 & 3.50 & -2.00 & 0.37 & 96.3 &  92.0 &  88.7 &  83.0 &  99.0 &  82.7\\

  6500 & 3.50 & -3.00 & 0.37 & 100.0 & 99.3 & 97.0 & 92.3 & 98.7 & 91.3\\[0.5ex]

  6500 & 4.50 & -1.00 & 0.41 &  0.0 &   0.0 &   0.0 &   0.3 & 30.7 &  41.0\\

  6500 & 4.50 & -2.00 & 0.40 & 97.3 &  96.0 &  90.0 &  79.3 & 98.7 &  86.3\\

  6500 & 4.50 & -3.00 & 0.40 & 100.0 & 100.0 & 97.7 & 96.0 & 100.0 & 90.0
  \\\tableline\tableline

\end{tabular}
\end{center}
\end{table}
}
 
A further test with {\em real\/} spectra of metal-poor stars confirms the high
level of completeness, though the number of involved spectra is admittedly
small. A cross-identification between 130 HES fields and the catalogue of
BPS92 yielded 8 turnoff-stars with $[$Fe/H$]<-3.0$. All of these have been
selected by the 3$\sigma$ criterium.
 
\subsection{Optimal Colour Range}
 
The optimal colour range for an efficient selection of the most metal-poor,
unevolved stars is $0.3 < (B-V) < 0.5$, since here no contamination of the
sample with late-type giants is expected.  The contamination with horizontal
branch stars will be small as well, since most of the colour range is overlapping
with the RR Lyrae gap, i.e. $0.2\lesssim (B-V) \lesssim 0.4$ (Walker 1998).
 
\section{A First Set of New Metal-Poor Stars From the HES}
 
In 1997 we selected a first set of candidates for metal-poor stars on HES
plates by automatic classification, i.e., with a {\it different} method than
described above. A learning sample of artificial, continous spectra was
used. Follow-up observations were carried out with EMMI attached to the
ESO-NTT at a spectral resolution of $\sim 5$\,{\AA}. Due to bad weather only
six stars could be observed with a sufficient $S/N$ ($\gtrsim 100$ at
H$\alpha$) to determine their stellar parameters $T_{\mbox{\tiny eff}}$,
$[$Fe/H$]$ and $\log g$ by spectral syntheses. The Balmer lines were used as a
temperature indicator, Mg\,{\sc i}~b as a gravity indicator, and the Fe lines as a
metallicity indicator. The results are given in Tab. \ref{mphsresults97oct}.
 
{\small
\begin{table}
  \caption{\label{mphsresults97oct}  Stellar parameters of new metal
    poor stars discovered by the Hamburg/ESO survey. The spectrum of
    HE~0008--3842 shows inconsistent temperatures between different Balmer
    lines. For this star we list the parameters of the best possible fit.}
  \begin{center}
    \begin{tabular}{lcccc}\tableline\tableline
      Name\rule{0.0mm}{2.5ex} & $B_J$ & $T_{\mbox{\scriptsize eff}}$
      & $\log g$ & $[$Fe/H$]$ \\\tableline
      HE~0131--2740 & 15.1 & $5600\pm 200$\,K & $3.5\pm 0.5$
      & $-2.1\pm 0.5$ \rule{0.0mm}{2.5ex}\\
      HE~0008--3842 & 14.4 & ($4700\pm 300$)\,K & ($1.5\pm 0.5$)
      & ($-2.5\pm 0.5$) \\
      HE~0119--4211 & 15.6 & $5300\pm 200$\,K & $2.0\pm 0.5$
      & $-2.3\pm 0.5$ \\
      HE~0442--5113 & 15.5 & $5900\pm 200$\,K & $3.7\pm 0.5$
      & $-1.9\pm 0.5$ \\
      HE~0350--4804 & 16.4 & $6100\pm 200$\,K & $4.5\pm 0.5$
      & $-2.0\pm 0.5$ \\
      HE~0448--3524 & 15.3 & $6100\pm 200$\,K & $3.8\pm 0.3$
      & $-0.4\pm 0.3$ \\\tableline\tableline
    \end{tabular}
  \end{center}
\end{table}
}
 
Five of the six analysed stars are metal-poor, while one has
$[\mbox{Fe/H}]=-0.4$.  Given the experience we have gained at present, this
star could have been easily rejected from the sample, because Mg\,{\sc i}~b is
visible in the objective-prism spectrum.
 
\section{Discussion}
 
The above results suggest that for an efficient selection of the most
metal-poor, unevolved stars we should restrict ourselves to objective-prism
spectra with $S/N>15$ in the colour range $0.3 < (B-V) < 0.5$.  By applying
the simulated selection function (Tab. \ref{selectionfunction}) to the
catalogue of BPS92, we estimate that at least $\sim$80\% of our candidates
will be unevolved stars; $\sim 40$\,\% of these having $[$Fe/H$]<-2.5$, and
every tenth star having $[$Fe/H$]<-3.0$.  A typical number of candidates is 50
per plate, and among them $\sim 10$ first-rank candidates.  Therefore, even
with the above restrictions applied, we will only be able to inspect our best
candidates by single-slit spectroscopic follow-up observations. The most
interesting of these will by investigated at high resolution using UVES and
the VLT for age determinations and abundance studies.

An improved efficiency of our survey with respect to the yield of the most
metal-poor stars is expected when automatic classification is used for the
candidate selection, as more spectral features will be involved. An especially
interesting feature is the G band of CH. Rossi et al. (1999, this volume) show
that the most carbon-enhanced stars from the HK survey are found among the
most metal-poor stars, so we expect that a selection of stars exhibiting
strong G bands, and neither Mg\,{\sc i}~b nor Ca\,{\sc ii}~K, will result
in a sample of carbon-enhanced {\em and\/} extremely metal-poor stars.
 
\acknowledgments N.C. was supported by DFG under grant Re~353/40-1 and travel
grant Ch~214/1-1. T.C.B. acknowledges support for this work from AST~95-29454
awarded by the NSF.


\end{document}